\documentclass{desyproc}
\begin{document}
%------------------------------------

\title{CP-violation in SUSY cascades at the LHC}

% for the author list please adhere to the format of one of the following
% three examples

% use the following for a single author
%
%\author{{\slshape Joe Smith}\\[1ex]
%DESY, Notketra{\ss}e 85, 22607 Hamburg, Germany }

% use the following for several authors
%
%\author{{\slshape Jean Meunier$^1$, Ruth Miller$^2$,
%    Gerd M\"uller$^3$\footnote{Speaker}, Joe Smith$^3$}\\[1ex]
%$^1$CERN, 1211 Gen\`eve 23, Switzerland\\
%$^2$Fermilab, P.O. Box 500, Batavia, IL 60510-0500, USA\\
%$^3$DESY, Notketra{\ss}e 85, 22607 Hamburg, Germany}

% use the following for an author speaking on behalf of a collaboration
%
\author{{\slshape Jamie Tattersall$^1$\footnote{Corresponding author, E-mail: jamie.tattersall@durham.ac.uk}}\,, Gudrid Moortgat-Pick$^2$, Krzysztof Rolbiecki$^1$  \\[1ex]
$^1${IPPP, University of Durham, Durham DH1 3LE, UK} \\
$^2${II. Institut fuer Theoret. Physik, University of Hamburg , Luruper Chaussee 149, D-22761 Hamburg, Germany}}

% please do not modify the following 5 lines
\contribID{xy}  % will be entered by the editors
\confID{1964}
\desyproc{DESY-PROC-2010-01}
\acronym{PLHC2010}
\doi            % will be entered by the editors

%\begin{flushright}
%  DCPT-10-128 \\
%  IPPP-10-64 \\
%  DESY-10-117 \\
%\end{flushright}

\maketitle

\begin{abstract}
 We study the potential to observe CP-violating effects in SUSY cascade decay chains at the LHC. Asymmetries composed by triple products of momenta of the final state particles are sensitive to CP-violating effects. Due to large boosts that dilute the asymmetries, these can be difficult to observe. Extending the methods of momentum reconstruction we show that the original size of these asymmetries may be measurable. A study is done at the hadronic level with backgrounds to estimate the expected sensitivity at the LHC.
\end{abstract}

\section{Introduction}

The search for Supersymmetry (SUSY) is one of the main goals of
present and future colliders since it is one of the best motivated
extensions of the Standard Model (SM). An important feature of SUSY
models is the possibility of incorporating new sources of CP
violation that are required to accommodate the baryon asymmetry of the universe. A
careful analysis of how to observe new CP-violating effects at the LHC will be required and in the following we discuss
an example in the Minimal Supersymmetric Standard Model.

CP-odd observables are the unambiguous way of discovering hints of
complex parameters in the underlying theory. One example of
such observables are CP-sensitive asymmetries based on the exploitation of triple product correlations of
momenta and/or spins of three final state particles with independent momentum, see also
\cite{Kittel:2009fg}. 

\begin{wrapfigure}{r}{0.45\textwidth}
%  \centering
  \begin{picture}(7,6.8)
 \put(1,0.3){\epsfig{file=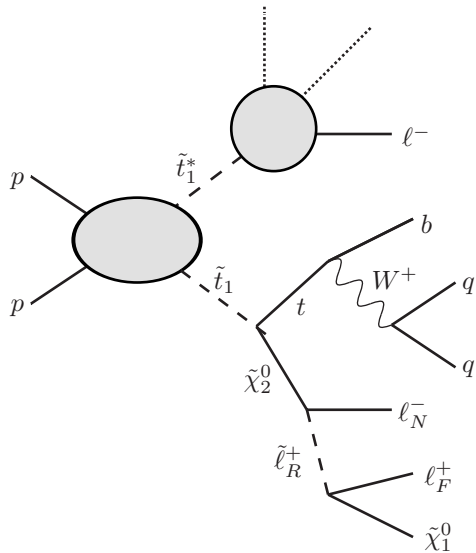,scale=0.5}} %(1,0.3)
    \put(0.9,5.1){$p$}
   \put(0.9,3.4){$p$}
   \put(3.6,3.7){$\tilde{t}_1$}
    \put(3.1,5.2){$\tilde{t}^*_1$}
     \put(6.1,5.6){$\ell^-$}
     \put(6.35,4.45){$b$}
     \put(4.7,3.35){$t$}
      \put(5.7,3.7){$W^+$}
     \put(4.0,2.4){$\tilde{\chi}^0_2$}
     \put(6.9,3.7){$q$}
     \put(6.9,2.6){$q$}
     \put(6.4,1.1){$\ell^+_F$}
       \put(6.4,.3){$\tilde{\chi}^0_1$}
   \put(4.4,1.3){$\tilde{\ell}^+_R$}
   \put(6.05,1.95){$\ell^-_N$}
 \end{picture}
   \vspace{-0.8cm}   
   \caption{\label{fig:FullDiagram} The process studied for momentum reconstruction.}
	\vspace{-1.98cm}
\end{wrapfigure}

Here we examine the production of $\tilde{t}_1\tilde{t}_1^*$ at the LHC with the following decay chain, Fig.~\ref{fig:FullDiagram},
\begin{equation}
 \tilde{t}_1 \to \tilde{\chi}^0_2 t, \qquad \tilde{\chi}^{0}_2 \to \tilde{\ell} \ell_N, \qquad \tilde{\ell} \to \tilde{\chi}^0_{1} \ell_F, \qquad t \to bjj. \label{eq:SimpDec}
\end{equation}

In this process \cite{MoortgatPick:2010new} the main source of CP-violation comes from the phase of the top trilinear coupling $A_t = |A_t| \mathrm{e}^{\mathrm{i} \phi_{A_t}}$. As an observable we choose the T$_N$-odd triple product of momenta of the final state particles,
\begin{equation}\label{eq:triple}
\mathcal{T}= \vec{p}_{\ell_N} \cdot (\vec{p}_{W} \times \vec{p}_{t})\; .
\end{equation}
Using this triple product one can construct a CP-odd asymmetry,
\begin{equation}\label{eq:asy}
\mathcal{A}_T = \frac{N_{\mathcal{T}_+}-N_{\mathcal{T}_-}}{N_{\mathcal{T}_+}+N_{\mathcal{T}_-}}\; ,
\end{equation}
where $N_{\mathcal{T}_+}$ ($N_{\mathcal{T}_-}$) are the numbers of events for which
$\mathcal{T}$ is positive (negative).

At the parton level, in the stop $\tilde{t}_1$ rest frame, the asymmetry can be as large as $15\%$, cf.\ Fig.~\ref{fig:PartonAsy}(a). However, particles produced at the LHC get large, undetermined boosts that are a consequence of the internal proton structure. Due to these boosts the asymmetry is strongly diluted as can be seen by comparing Fig.~\ref{fig:PartonAsy}(a) and Fig.~\ref{fig:PartonAsy}(b). This makes the observation and analysis of CP-violating effects very difficult at the LHC. For further discussion of these effects and other studies of CP-violation in stop decays at the LHC see~\cite{Ellis:2008hq}.

We show that a very useful tool in such an analysis is the reconstruction of momenta of all the particles involved in the process, including those escaping detection ($\tilde{\chi}^0_1$). Using this technique one can recover the large asymmetry present at the parton level by boosting back into the rest frame of the stop. Furthermore, we can heavily suppress both standard model and SUSY backgrounds using reconstruction. Therefore, we greatly increase the discovery potential and here we present the first hadronic study of momentum reconstruction in relation to CP-violation. The technique of momentum reconstruction for CP-violating observables was first presented in~\cite{MoortgatPick:2009jy}.

\section{CP-violation in the laboratory frame}

\begin{figure}[ht!]
\begin{picture}(16,8)
 \put(6,2.5){$\phi_{A_t}/\pi$}
 \put(0,8){\epsfig{file=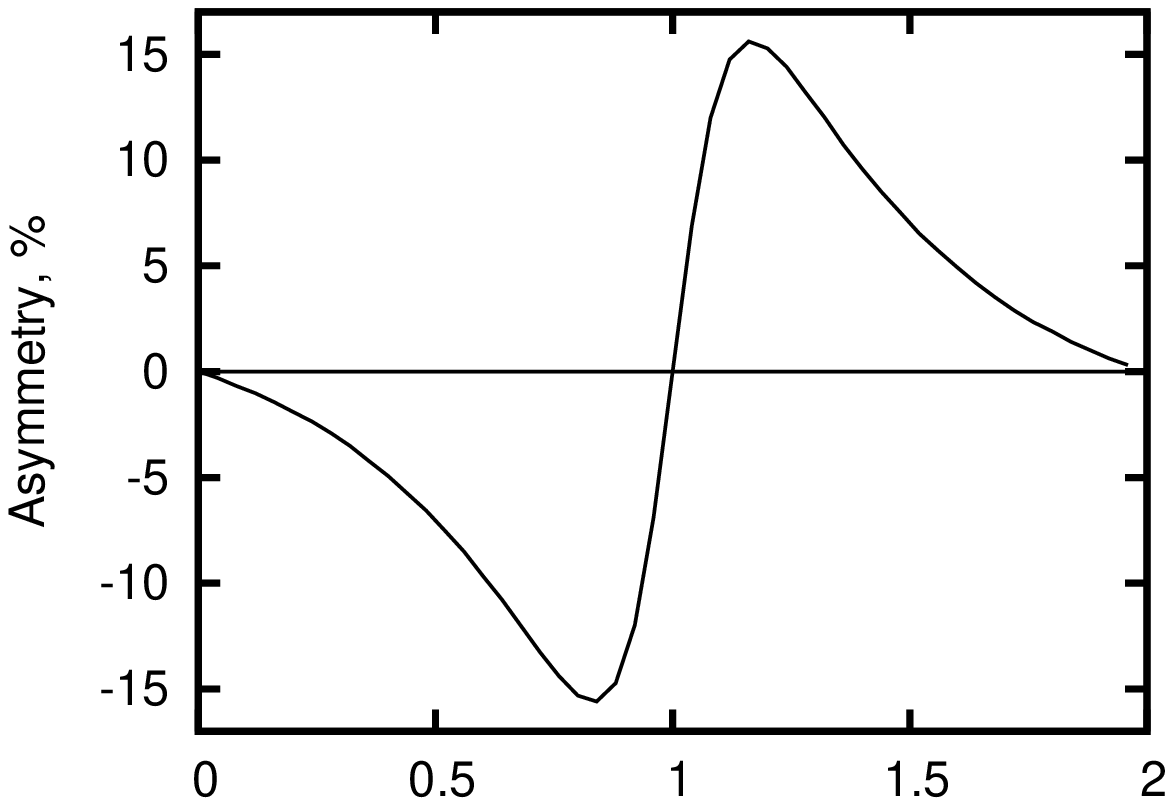,scale=0.6,angle=270}}
  \put(0,7.5){(a)}
  \put(14,2.5){$\phi_{A_t}/\pi$}
  \put(8,8){\epsfig{file=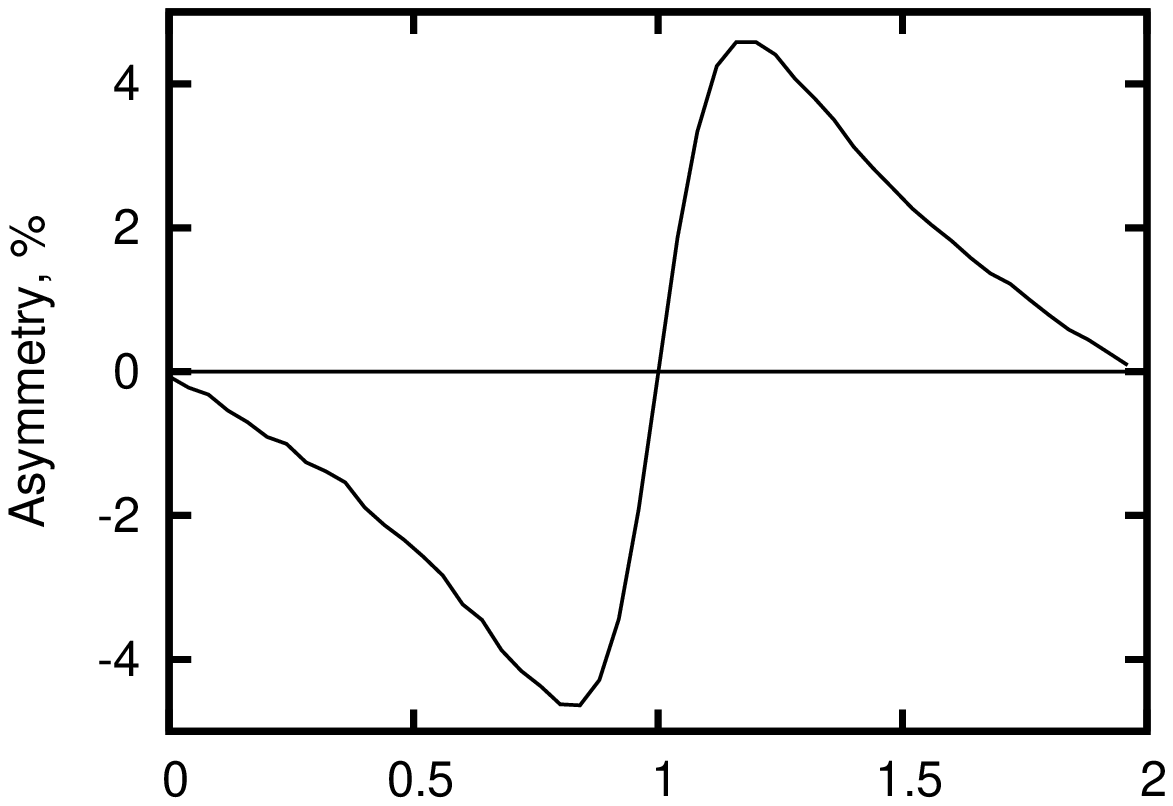,scale=0.6,angle=270}}
  \put(13,8.0){$\sqrt{s}=14\mathrm{TeV}$}
   \put(8,7.5){(b)}
\end{picture}
\vspace{-3cm}
\caption{\label{fig:PartonAsy} The asymmetry $\mathcal{A}_{\mathrm{T}}$ as a function of $\phi_{A_t}$, Eq.~(\ref{eq:asy}). (a) in the rest frame of $\tilde{t}_1$, (b) in the laboratory frame at the LHC at 14TeV.}
\end{figure}

First we study the behaviour of the asymmetry after the inclusion of parton distribution functions (PDFs). Our observable, Eq.~\eqref{eq:asy}, is significantly reduced due to boosts compared with the asymmetry in the stop rest frame, where it is maximal, see Fig.~\ref{fig:PartonAsy}(a). This is because a boosted frame can make the momentum vector of the lepton appear to come from the opposite side of the plane formed by $W$ and $t$, hence changing the sign of the triple product, Eq.~(\ref{eq:triple}). Inclusion of PDFs reduce the asymmetry by about factor of 4 in our case. The maximum asymmetry is about $|\mathcal{A}_T| \simeq 4.5\%$ and if we use this asymmetry at the LHC it would be of limited statistical significance. 

\section{CP-violation with momentum reconstruction}

In order to overcome the dilution factor due to PDFs, we investigate the possibility of reconstructing the
momenta of the invisible particles ($\tilde{\chi}^0_1$) in the process on an event by event basis~\cite{MoortgatPick:2009jy,Kawagoe:2004rz}. We perform the reconstruction at the hadronic level to verify the viability of the technique to the LHC.

For the decay chain of interest we can reconstruct the four unknown components of the $\tilde{\chi}^0_1$ momentum assuming me know the masses of the four particles involved in the cascade decay. We write down the four on-shell mass conditions,
\begin{eqnarray}
   m_{\tilde{\chi}^0_1}&=&(P_{\tilde{\chi}^0_1})^2, \label{eq:mneu1} \\  
  m_{\tilde{\ell}}&=&(P_{\tilde{\chi}^0_1}+P_{\ell^+_F})^2,  \\
  m_{\tilde{\chi}^0_2}&=&(P_{\tilde{\chi}^0_1}+P_{\ell^+_F}+P_{\ell^+_N})^2,  \\
 m_{\tilde{t}_1}&=&(P_{\tilde{\chi}^0_1}+P_{\ell^+_F}+P_{\ell^+_N}+P_t)^2, \label{eq:mstop1} 
\end{eqnarray}
and solve the system.

Once we have the $\tilde{\chi}^0_1$ momentum we can trivially find the momentum of any other particle in the cascade decay. We can therefore find the momentum of the $\tilde{t}_1$ and boost all final state particles into this frame to recover the full asymmetry.

There is a complication in finding the $\tilde{\chi}^0_1$ momentum because if we solve Eq.~(\ref{eq:mneu1}-\ref{eq:mstop1}) we see that we are left with a quadratic in $(P_{\tilde{\chi}^0_1})^2$. Consequently, we will have two viable solutions for the $\tilde{\chi}^0_1$ momentum but we cannot know which is correct. Since we do not have any additional constraints to pick the correct solution, we calculate the $\tilde{t}_1$ rest frame for both. However we only count those events that give the same sign for the triple product, Eq.~(\ref{eq:triple}). This guarantees that we take the correct sign for the triple product for the calculation of the asymmetry. 

In addition, this method also significantly reduces the combinatorial background from wrong lepton or jet identification and both standard model and SUSY backgrounds \cite{MoortgatPick:2010new}. For example, we need to correctly identify the near and far lepton in the cascade decay, Eq.~\ref{eq:SimpDec}, and our method is to try to perform reconstruction with both lepton assignments. Firstly, only a small subset of events with the wrong assignment will give real solutions for the $\tilde{\chi}^0_1$ momentum. Secondly, if the wrong assignment does satisfy the kinematical conditions, we only accept events where the sign of all triple products coincide. As one solution will have the true assignment we know the sign of the triple product will be correct and thus we will kill the combinatorial background.

\begin{table}[ht!] \renewcommand{\arraystretch}{1.3}
\begin{center}
\begin{tabular}{|c||c|c|c|c|c|}\hline
Parameter & $m_0$ & $m_{1/2}$ &  $\tan\beta$ & sign($\mu$) &  $A_0$ \\ \hline\hline
Value & 65        & 210       &   5          & +           & 0     \\ \hline
\end{tabular}
\caption{mSUGRA benchmark scenario \label{tab:Scenario}}
\end{center}
\end{table}
\vspace{-0.5cm}
To estimate the viability of this measurement at the LHC we study a mSUGRA scenario (Tab.\ref{tab:Scenario}) and produce fully hadronic events with a jet finder applied. We also apply realistic cuts for the LHC, implement experimental efficiencies (e.g. $b$-tagging). In addition, we require a top within a mass window to be reconstructed and include the most important standard model backgrounds. Within this scenario we find that the LHC should have sensitivity at 3-$\sigma$ with 500 fb$^{-1}$ for $0.5\pi < \phi_{A_t} < 0.9\pi$.

\begin{figure}[ht!]
\vspace{1cm}
\begin{picture}(16,7)
   \put(1,2.2){\epsfig{file=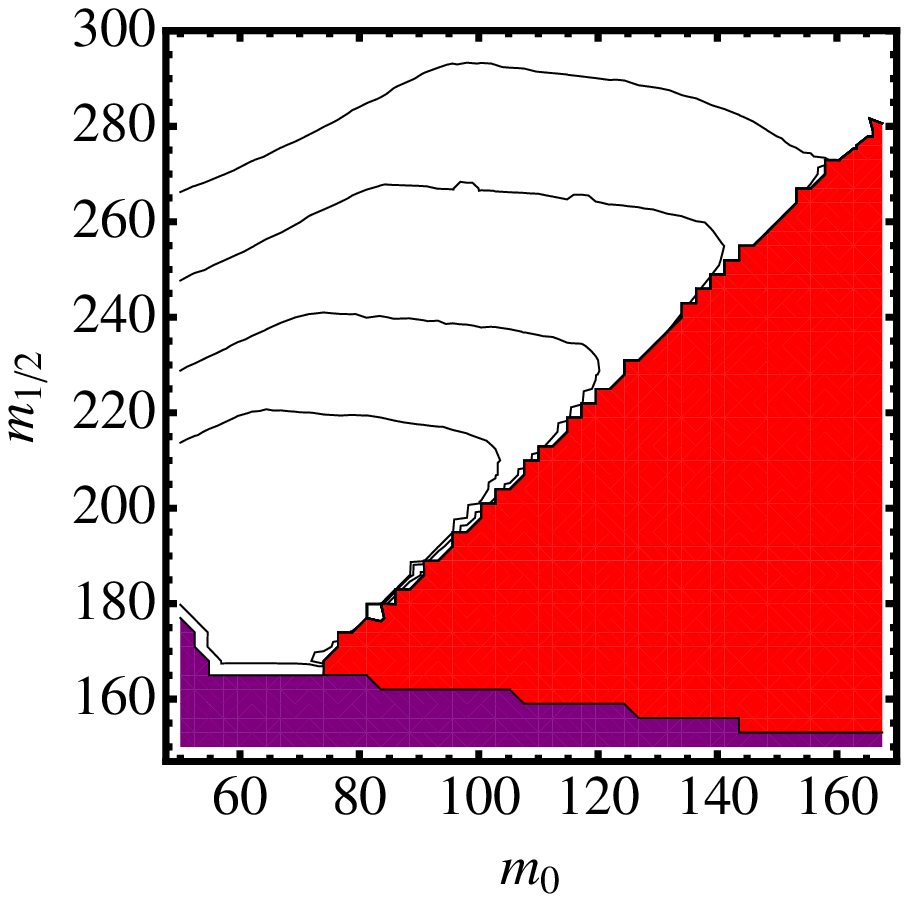,scale=0.6}}
      \put(2.7,4.9){\tiny{300 fb$^{-1}$}}
      \put(2.7,5.5){\tiny{500 fb$^{-1}$}}
    \put(2.7,6.2){\tiny{1 ab$^{-1}$}}
    \put(2.7,6.7){\tiny{2 ab$^{-1}$}}
	\put(3.9,7.8){$3\sigma$-observation}
    \put(0,7.5){(a)}

  \put(8.5,2.2){\epsfig{file=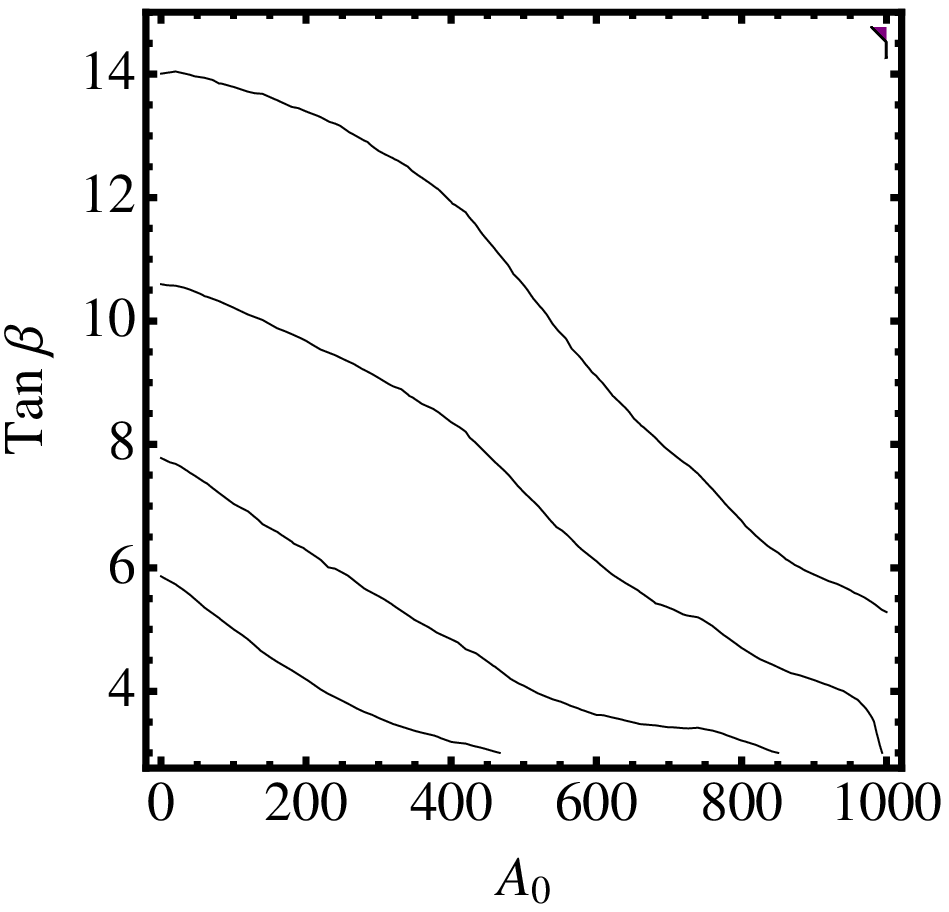,scale=0.6}}
      \put(9.6,3.2){\tiny{300 fb$^{-1}$}}
      \put(9.9,3.9){\tiny{500 fb$^{-1}$}}
    \put(10.3,5.0){\tiny{1 ab$^{-1}$}}
    \put(10.6,6.1){\tiny{2 ab$^{-1}$}}
	\put(11.4,7.8){$3\sigma$-observation}
   \put(8,7.5){(b)}
\end{picture}
\vspace{-3cm}
\caption{\label{fig:NoCutsM0} Minimum luminosity required for 3$\sigma$-discovery of CP-violation in $\tilde{t}_1\tilde{t}_1$ production at the LHC at 14 TeV. Purple (dark) area is ruled out by LEP direct detection and red (light) area has no two body decay $\tilde{\chi}^0_2 \to \tilde{\ell}^{\pm} \ell^{\mp}$. (a) $m_0, m_{1/2}$ plane, (b) tan$\beta, A_0$ plane.}
\end{figure} 

In Figs.~\ref{fig:NoCutsM0}(a,b) we see the effect of varying the mSUGRA parameters on the minimum luminosity required for a 3-$\sigma$ observation at the LHC, assuming that the parton level asymmetry, $|\mathcal{A}_T| = 15\%$. We see that as $m_{1/2}$ is increased, Fig.~\ref{fig:NoCutsM0}(a), we require more luminosity due to an increased $\tilde{t}_1$ mass and hence smaller production cross section. An increase in both tan$\beta$ and $A_0$ also produce a similar increase in the luminosity required for discovery. An increase in tan$\beta$ reduces the $\tilde{\chi}^0_2 \to \tilde{\ell}^{\pm} \ell^{\mp}$ branching ratio while an increase in $A_0$ produces a smaller $A_t$ that consequently reduces our sensitivity to the phase.

\section{Acknowledgements}

KR is supported by the EU Network MRTN-CT-2006-035505 (HEPTools). JT
is supported by the UK Science and Technology Facilities Council
(STFC).

% ****************************************************************************
% BIBLIOGRAPHY AREA
% ****************************************************************************

% please do not change the following line
\begin{footnotesize}

% please do not change the following line
\end{footnotesize}

% ****************************************************************************
% END OF BIBLIOGRAPHY AREA
% ****************************************************************************


\begin{thebibliography}{99}

%------- replace following references ;-)

\bibitem{Kittel:2009fg}
  O.~Kittel,
  %``SUSY CP phases and asymmetries at colliders,''
  [arXiv:0904.3241 [hep-ph]].
  %%CITATION = 0904.3241;%%

\bibitem{MoortgatPick:2010new}
  G.~Moortgat-Pick, K.~Rolbiecki and J.~Tattersall,
  %``Momentum reconstruction at the LHC for probing CP-violation in the stop sector,''
   [arXiv:1008.2206 [hep-ph]]

\bibitem{Ellis:2008hq}
  J.~Ellis, F.~Moortgat, G.~Moortgat-Pick, J.~M.~Smillie and J.~Tattersall,
  %``Measurement of CP Violation in Stop Cascade Decays at the LHC,''
  {\it Eur.\ Phys.\ J.\ }{\bf C60}, 633 (2009)
  [arXiv:0809.1607 [hep-ph]].
  %%CITATION = EPHJA,C60,633;%%
  P.~Langacker, G.~Paz, L.~T.~Wang and I.~Yavin,
  %``A T-odd observable sensitive to CP violating phases in squark decay,''
  JHEP {\bf 0707} (2007) 055
  [arXiv:hep-ph/0702068].
  %%CITATION = JHEPA,0707,055;%%

\bibitem{MoortgatPick:2009jy}
  G.~Moortgat-Pick, K.~Rolbiecki, J.~Tattersall and P.~Wienemann,
  %``Probing CP Violation with and without Momentum Reconstruction at the LHC,''
  JHEP {\bf 1001}, 004 (2010)
  [arXiv:0908.2631 [hep-ph]].
  %%CITATION = ARXIV:0908.2631;%%

\bibitem{Kawagoe:2004rz}
  K.~Kawagoe, M.~M.~Nojiri and G.~Polesello,
  %``A new SUSY mass reconstruction method at the CERN LHC,''
  {\it Phys.\ Rev.\ }{\bf D71}, 035008 (2005)
  [arXiv:hep-ph/0410160];\\
  %%CITATION = PHRVA,D71,035008;%%
  H.~C.~Cheng, D.~Engelhardt, J.~F.~Gunion, Z.~Han and B.~McElrath,
  %``Accurate Mass Determinations in Decay Chains with Missing Energy,''
  {\it Phys.\ Rev.\ Lett.\ }{\bf 100}, 252001 (2008)
  [arXiv:0802.4290 [hep-ph]].
  %%CITATION = PRLTA,100,252001;%%

\end{thebibliography}
\end{document}